\documentclass[newstyle,twocolumn,proceedings]{rmaa}

\usepackage{rmaacite}

\renewcommand{\P}[1]{%
\ifnum#1=1\hbox{OW~168--326E}\fi
\ifnum#1=2\hbox{OW~167--317}\fi
\ifnum#1=3\hbox{OW~163--317}\fi
\ifnum#1=5\hbox{OW~158--323}\fi
\ifnum#1=0\hbox{OW~171--334}\fi}
\makeatletter
\title{ULIRGs: Origin and Evolution\altaffilmark{1}}
\author{I. M\'arquez and J. Masegosa
  \affil{Instituto de Astrof\'{\i}sica de Andaluc\'{\i}a (CSIC), Granada
(Spain)} }
\altaffiltext{1}{The data presented here have been taken using ALFOSC, 
which is owned by the Instituto de Astrof\'{\i}sica de Andaluc\'{\i}a (IAA) and operated
at the Nordic Optical Telescope under agreement between IAA and the NBIfAFG 
of the Astronomical Observatory of Copenhagen}

\fulladdresses{
\item I. M\'arquez and J. Masegosa: Instituto de Astrof\'{\i}sica de
Andaluc\'{\i}a (CSIC), Apdo 3004, 18080 Granada (Spain) (isabel@iaa.es,
pepa@iaa.es).}

\shortauthor{M\'arquez \& Masegosa}
\shorttitle{ULIRGs: Origin and Evolution}


\abstract{
We present our preliminary results on the study of the high luminosity
ULIRGs (L$_{FIR} >$ 10$^{12.3}$L\sun) in the 1Jy sample. Deep, high resolution optical R images with
7 arcmin $\times$ 7 arcmin field of view for 2/3 of the whole sample,
show that they are generally found in very rich environments, they
seem to be late stage merger products (long tidal tails are found in
only 3 systems) and all of them show multiple nuclei (double nuclei in
11: 2HII, 4 LINERs, 4 Sy2 and 1 Sy1). GTC first light instruments are
very well suited for addressing the question of the energy source of
ULIRGs, together with a more refined view of the proposed evolutionary
sequence.
}
\resumen{Presentamos los resultados preliminares sobre un estudio de la banda 
de alta luminosidad (L$_{FIR} >$ 10$^{12.3}$L\sun) de las ULIRGs de la muestra de 1Jy. Im\'agenes profundas, 
de alta resoluci\'on y campo de 7 $\times$ 7 minutos para 2/3 de la muestra 
indican que las ULIRGs se encuentran generalmente en entornos muy ricos, que 
se parecen a lo que se espera en los \'ultimos estadios de la fusi\'on de 
galaxias (encontramos colas de marea largas en s\'olo 3 sistemas) y todas 
ellas tienen n\'ucleos m\'ultiples (son dobles en 11 ULIRGs: 2 HII, 4 LINERs, 
4 Sy2 y 1 Sy1). Los instrumentos de primera luz previstos para GTC son 
especialmente indicados para el estudio de la fuente de energ\'{\i}a, y 
permitir\'an tambi\'en obtener una visi\'on m\'as clara de la secuencia 
evolutiva propuesta. }

\listofauthors{I.~M\'arquez \& J.~Masegosa}
\indexauthor{M\'arquez, I.}
\indexauthor{Masegosa, J.}

\begin{document}

\maketitle

\section{Introduction}
\label{sec:intro}
The most conspicuous finding by IRAS \cite{Soifer87} was the
discovery of a new population of galaxies emitting at FIR
wavelengths. Their energies are comparable to those of the most
luminous quasars. Such similar power between QSO and luminous infrared
galaxies could be explained considering the latter as dust-enshrouded
QSOs. The inmediate question to answer is the nature of the energy
source.

The first IRAS data concerned a sample of 9 galaxies, optically very
faint (L$_{IR}$/L$_B$ $>$ 20). Only two of those galaxies had been already
classified in the optical range, both with very perturbed
morphologies: Arp 220 and NGC 6240. The most likely explanation for
the large observed FIR luminosities ($\approx$10$^{12}$L\sun) is that
they should host very strong UV emitting sources, very deeply obscured
by large amounts of dust; the dust would be heated by the UV radiation
and re-radiate at FIR. During the late 80's and the beggining of 90's
the question wether a black hole or a starburst is the nuclear source
was posed. Rowan-Robinson's models for their SED's showed that,
whereas both cirrus and starburst components were always needed, the
active nucleus was not invoqued in all cases.

Optical spectra showed that most of these galaxies ressembled better
those of LINERs than Seyferts 1 or 2. The characterization of a
complete sample of ULIRGs had to await until 1995:
\scite{Veilleux95,Veilleux99}  showed that the
percentage of HII-like nuclei, LINERs and Seyferts changes as a
function of the IR luminosity. For low L$_{FIR}$ most of the nuclei are
starburst-like; up to 60\% of the most luminous ULIRGs are
Seyfert-like, only 20\% of them being HII-like. Interestingly, the
percentage of LINERs (30\%) remains constant in spite of their L$_{FIR}$
level. An evolutionary scheme appers to explain this dependence on
L$_{FIR}$ of the nature of the energy source.

\section{An evolutionary sequence} 
\label{sec:sequence}
Strongly interacting systems are highly frequent among ULIRGs, and
their properties are therefore studied in connection with the effects
of merging processes. \scite{Sanders88} suggested that ULIRGs would
evolve in luminosity as precursors of optically selected
quasars. Following this scenario, massive star formation induced by
stong gravitational interaction would be the first energy source;
later in the merger sequence more and more star formation is produced
and eventually very compact stellar clusters are formed, that could
give rise to the formation of a massive black hole (BH);
the BH should be fed up with material from the circumnuclear regions,
where starbursting processes would take place and consequently an
obscured QSO could be observed.

The morphologies of ULIRGS have been classified \cite{Surace98,Surace00,Surace99} as:
I. Pre-contact; II. First contact without tidal tails; Pre-merger with
tidal tails and double nuclei separated (a) more than 10 kpc and (b)
less than 10 kpc; IV. Merger with long tidal tails and a single (a)
diffuse and (b) compact nucleus; V. Final stage merger, no tidal tails
and strong central perturbations.

Following this classification, \scite{Veilleux01} has found that the
percentage of HII-like, LINER and Seyfert (S1, S2) nuclei is: III:
40\% HII, 40\% LINER, 20\% Sy2; IV: 25\% HII, 38\% LINER, 43\% Sy2,
15\% Sy1; V: 25\% HII, 35\% LINER, 40\% Sy2, 15\% Sy1. That is to say, more
advanced mergers host more AGNs (LINERs are constant).

\section{Our aims and preliminary results}
\label{sec:results}
We plan to characterize the environmental status together with the
merger stage, paying attention not only to the very central regions
but also to large scale structures that would allow a better
determination of the ``age'' in the expected merger sequence. To do
so, high resolution and large field of view optical imaging is being
obtained. First, we are in the process of getting R-band images of the
whole high luminosity strip (L$_{FIR} >$ 10$^{12.3}$L\sun, 30
galaxies). Among the 19 galaxies observed up to now (2 HII-like, 8
LINERs, 4 Sy2, 5 Sy 1, see the figure for one example) we obtain the
following, preliminary conclusions: (1) They are generally found in
very rich environments; (2) they seem to be late stage merger products
(long tidal tails are found in only 3 systems); (3) all of them show
multiple nuclei (double nuclei in 11: 2HII, 4 LINERs, 4 Sy2 and 1
Sy1).
  
\begin{figure}
  \includegraphics[width=\columnwidth]{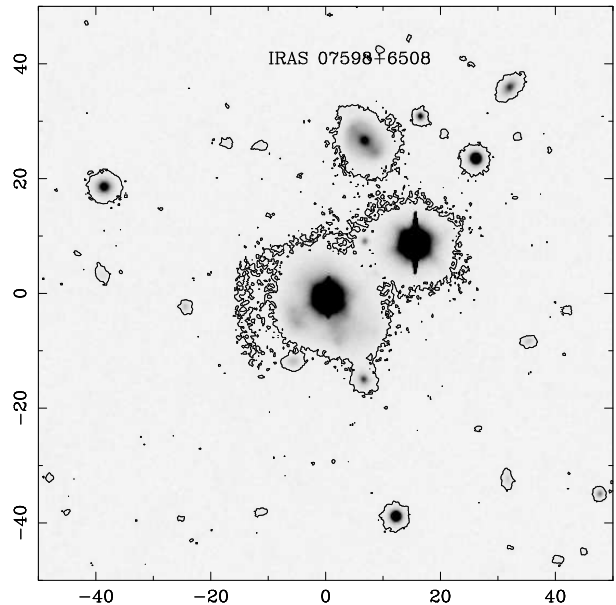} 
\caption{R band image of IRAS 07598+6508 (FWHM = 0.8 arcsec).   
The contour corresponds to $\mu_R$ = 25.36 mag/arcsec$^2$. Tics are in arcseconds from the source.}  
\label{figure}
\end{figure}

\section{ULIRGs with GTC}
\label{sec:gtc}
Both the resolution and collecting power of a 10m-class telescope will
allow to study the properties of ULIRGs from our nearby Universe to
higher redshifts. The question of the nature of the energy source will
be obviously better addressed, together with a more refined view of
the proposed evolutionary sequence from one type of object to the
other, and the study of the nature of the precursors of today's
ULIRGs. First light instruments are very well suited for these
purposes, both OSIRIS, to get optical to NIR imaging or spectroscopy,
and CANARICAM, to get the valuable information provided at MIR
wavelengths (note that ULIRGs are one of the scientific cases for
CANARICAM). The determination of the nature of the energy source will
be completed with the advent of EMIR.

\acknowledgements We acknowledge financial support from the Spanish
Ministerio de Ciencia y Tecnolog\'{\i}a through the programs PB98-0521 
and AYA2001-2089.


\end{document}